\documentclass[twocolumn]{aa} 

\usepackage{graphicx}
\newcommand{\peijin}[1]{#1}
\newcommand{\zpeijin}[1]{#1}

\usepackage{stfloats}
\usepackage{txfonts}
\usepackage{hyperref}
\hypersetup{
    colorlinks=true,
    linkcolor=blue,
    filecolor=magenta,      
    urlcolor=cyan,
    pdftitle={Overleaf Example},
    pdfpagemode=FullScreen,
    }
\begin{document}

   \title{Imaging spectroscopy of spectral bump in a type II radio burst}
\titlerunning{Spectra bump in Type II}

   \author{Peijin Zhang
          \inst{1,2,3}
          \and
          Diana E. Morosan\inst{1,4}
          \and
          Pietro Zucca \inst{5}
          Sanna Normo\inst{4}
          \and
          Bartosz Dabrowski\inst{6}  
          \and
          Andrzej Krankowski \inst{6}
          \and
          Christian Vocks \inst{7}
}
   \institute{  Department of Physics, University of Helsinki, P.O. Box 64, FI-00014 Helsinki, Finland 
   \and
            Center for Solar-Terrestrial Research, New Jersey Institute of Technology, Newark, NJ, USA
        \email{peijin.zhang@njit.edu}
        \and
         Cooperative Programs for the 
Advancement of Earth System Science, University Corporation for Atmospheric Research, Boulder, CO, USA
         \and
         Department of Physics and Astronomy, University of Turku, FI-20500 Turku, Finland
         \email{diana.morosan@utu.fi}
         \and
         ASTRON - the Netherlands Institute for Radio Astronomy, Oude Hoogeveensedijk 4, 7991 PD Dwingeloo, the Netherlands
         \and
         Space Radio-Diagnostics Research Centre, University of Warmia and
Mazury, R. Prawochenskiego 9, Olsztyn, 10-719, Poland.
\and
Leibniz-Institut für Astrophysik Potsdam (AIP)
An der Sternwarte 16, 14482 Potsdam, Germany
}

   \date{Received \today,; accepted \today}

 
  \abstract
{Observations of solar type II radio bursts provide a unique opportunity to analyze the non-thermal electrons accelerated by coronal shocks and also to diagnose the plasma density distribution in the corona. However, there are very rare high-frequency resolution interferometric observations for type II radio bursts that are capable of tracking these electrons.}
{Recently, more spatially resolved high-resolution observations of type II radio bursts have been recorded with the Low-Frequency Array (LOFAR). Using these observations, we aim to track the location of a type II radio burst that experiences a sudden spectral bump.}
{Here, we present the first radio imaging observations for a type II burst with a spectral bump. We measure the variation in source location and frequency drift of the type II burst, and deduct the density distribution along its propagation direction.}
{We identified a type II burst that experiences a sudden spectral bump in its frequency-time profile. The overall frequency drift rate is 0.06~MHz/s and it corresponds to an estimated speed of 295~km/s. The projected speed of the radio source obtained from imaging is 380~km/s towards the east direction. At the spectral bump, a deviation in the source locations of the type II split bands is observed. The band separation increases significantly in the north-south direction.}
{
The spectral bump shows an 8~MHz deviation at 60~MHz which corresponds to a 25\% decrease in the plasma density. The estimated crossing distance during the spectrum bump in type II is 29~Mm suggesting that this density variation occurs in a confined area.
This indicates that the shock most likely encounters the upper extent of a coronal hole.}
   \maketitle
%

\section{Introduction}

Solar radio bursts present a crucial diagnostic of the underlying acceleration processes and conditions in the solar atmosphere. These emissions can provide information on the energy release process in solar activities and also provide diagnostics on the background plasma of the solar corona \citep{wild1950typeII}.
Solar radio bursts in the low-frequency radio range are categorized into five main types (type I-V) based on the dynamic spectrum morphology. These five types of radio emission correspond to different physical processes in the solar corona. Type II radio bursts are generally accepted to be associated with coronal shocks \citep[e.g.,][]{mann2005electron,morosan2023type, ramesh2023ApJsolarcoronadens}.

Type II solar radio bursts are interpreted as the signature of shock-accelerated electron beams generating Langmuir waves close to the local plasma frequency ($f_p$), and then converted into radio emission at fundamental (F) and harmonic (H) frequency of $f_p$ \citep[e.g.,][]{melrose1980emission,NelsonMelrose1985srph, Cairns2003SSRtypeII}. 
The overall shape of a type II radio burst slowly drifts  from high to low frequencies in the dynamic spectrum. Within this drifting feature, there are lots of other fine-structured features, reflecting the background density variation and/or electron beam properties \citep[e.g.,][]{Cairns1987SoPh, Jasmina2020ApJfine, Diana2019NatAs, morosan2024, zhang2024}.
One such interesting feature is the short-term ($\sim$minute) significant variation or jumps from the general direction of the drifting lane ($\Delta f/f  \sim 10\%$, where $f$ is the emission frequency). Previous observations of a spectral jump or bump \citep{AKovaltype2bump2021ApJ,fengBumpTypeII2013ApJ} show that the frequency is offset towards a higher frequency and this is attributed to the interaction of the shock with high-density structures (e.g., coronal streamers). Until now, there has been no radio imaging of spectral bumps to show the variation in location of the radio source when it experiences such a bump.

\begin{figure*}[ht]
    \centering
    \includegraphics[width=0.95\textwidth]{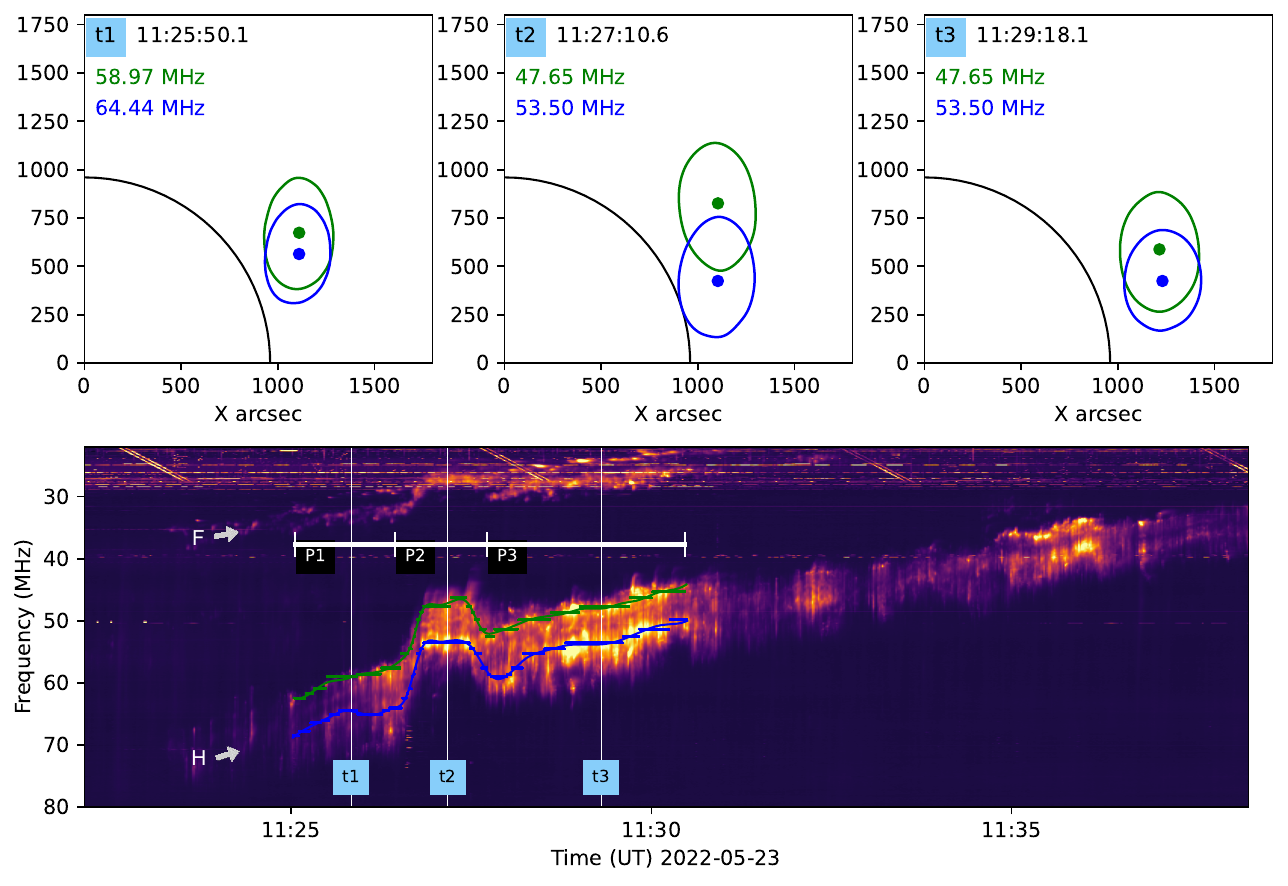}
    \caption{Dynamic spectrum and imaging of three time-points in the type II radio burst.
    Blue represents the higher frequency lane. The upper 3 panels show the source size and location at these three time-slots, the dot marks the peak location and the solid line marks the half-maximum contour. The solid black lines in the upper panel represent the solar disk.
    The lower panel is the dynamic spectrum, the solid green and blue line marks the upper and lower lane of the type II. White horizontal marks the 3 phases of the burst. P1: before bump, P2: bump, P3: after bump. The imaging of the three time-slots (t1,t2,t3) is presented in the upper 3 panels. \zpeijin{F and H mark the fundamental and harmonic components in the dynamic spectrum.} }
    \label{fig:main-ds}
\end{figure*}

In the quest to understand these phenomena better, radio imaging has emerged as an invaluable tool. Unlike traditional spectrographs which provide only spectral information, radio imaging can spatially resolve the source regions of these emissions. This spatial resolution is pivotal in tracing the origins, paths, and interactions of these radio waves with the ambient solar atmosphere. Recent advancements in radio telescope technology have markedly improved our understanding of type II radio burst properties, offering significant insights into their evolution. For example, \cite{Diana2019NatAs} used tied-array beam imaging observations with LOFAR to image herringbone bursts and identified the presence of shock-accelerated electron beams at multiple regions around the CME shock.
\cite{Maguire2021lofar} used LOFAR interferometry imaging combined with EUV observations, showed that a type II radio source is located above a solar jet, which suggests the type II was generated by a jet-driven piston shock, instead of a CME-driven shock wave.
In another study with the Murchison Widefield Array (MWA), \cite{bhunia2022imagingspectroscopy} performed radio imaging for a band-split type II burst at multiple frequencies, and found that the radio source is located at multiple locations close to the shock.

Despite the potential of radio imaging, there has been a distinct lack of high-quality imaging studies specifically targeted at type II radio bursts using modern interferometers. The Low Frequency Array (LOFAR) \cite{van2013lofar} is a radio telescope array with antenna stations spreading across Europe. It operates at the lowest frequencies that can be observed from Earth, ranging from 10 to 240 MHz. The imaging capability provides the possibility of spatially resolving the radio signals from the coronal shock.


In this paper, we aim to bridge this gap by presenting a comprehensive study using advanced interferometry imaging techniques. Section 2 presents the observation detail, including data processing and event overview. In Section 3, we show the analysis of the type II bump,  and
in Section 4 discussion and conclusion.

\section{Observation}

We present imaging and spectroscopy observations of a type II radio burst observed on 2022-05-23 at 11:25UT (see Fig \ref{fig:main-ds}), with an identifier of \texttt{LT16\_001} in the LOFAR long-term archive.
The observation is done with 31 Low Band Antenna (LBA) stations from LOFAR, of which there are 23 core-stations, and 8 remote stations, the longest available baseline being 48~km. These 31 stations generate 496 station baselines.
We used 60 frequency channels from 20 to 80MHz, with a 195.3~kHz channel width.
The time resolution of imaging is 0.671s.We used two simultaneous beams pointing at the Sun and a calibrator (Cassiopeia-A). In the data processing steps, the calibrator is used to correct the phase and amplitude variations of the Sun observations. The dynamic spectrum is recorded by a stand-alone core station (CS032) with a time resolution of 10.5 milliseconds and a frequency resolution of 12.5kHz. 

\subsection{Data processing}

The dynamic spectrum is pre-processed with RFI-flagging tool for solar and spaceweather spectrum: \texttt{ConvRFI} \footnote{ConvRFI \href{https://github.com/peijin94/ConvRFI}{https://github.com/peijin94/ConvRFI}} \citep{zhang2023MNRASConvRFI}.

The interferometry data processing involves the following steps: 

\begin{enumerate}
    \item \textbf{Gain Calibration:} For the calibrator observation,  we compute the phase and amplitude offsets with a flux density model of the calibrator as a reference. This step will generate a gain solution of phase and amplitude for each baseline and subband. The computation of gaincal is performed with \texttt{DP3} (The Default Pre-Processing Pipeline) \footnote{DP3 \href{https://github.com/lofar-astron/DP3}{https://github.com/lofar-astron/DP3}} \cite{2018asclsoft04003V}.
    \item \textbf{Antenna Inspection:} Review the phase and amplitude plots for each antenna and flag corrupted data.
    \item \textbf{Applying the calibration:} The gain solution is applied to the Sun observation to correct phase and amplitude. The computation of Applycal is performed with \texttt{DP3}.
    \item \textbf{Imaging:} We performed a 2D Fourier Transform of the visibility-map. Then, we executed the w-stacking CLEAN algorithm to deconvolute the point-spread-function (PSF) from the image. The imaging step is performed with \texttt{wsclean} \footnote{wsclean \href{https://gitlab.com/aroffringa/wsclean}{https://gitlab.com/aroffringa/wsclean}} \citep{offringaWsclean2014}.
    \item \textbf{Post-processing:} We converted coordinates to helio-projective and the units to brightness temperature using\texttt{lofarSun} \footnote{lofarSun \href{https://github.com/peijin94/LOFAR-Sun-tools}{https://github.com/peijin94/LOFAR-Sun-tools}} \citep{Zhang2022lofarsun}.
\end{enumerate}

To characterize the spectroscopic features of the type II burst, we extracted the time and frequency evolution of each split-band. We used a "smoothed line of local maximum" for the upper and lower bands (shown as green and blue lines in Figure \ref{fig:main-ds}). Interferometric imaging is performed along the two bands picking the closest available time-frequency points (shown as green and blue dots in Figure \ref{fig:main-ds}).

\subsection{The type II spectral bump}

In this Type II solar radio burst, shown in the dynamic spectrum (Fig. \ref{fig:main-ds}).
\zpeijin{The burst shows a fundamental-harmonic pair structure, with the fundamental starting near 35~MHz and the harmonic starting close to 70~MHz at $\sim$11:25~UT.
In the harmonic component of the} type II a notable anomaly appears as a distinctive `bump' within the frequency drift evolution. Here, the signal shifts towards lower frequencies by approximately 5-10 MHz, persisting for a duration of one minute before returning to the overall frequency drift evolution.
The analysis in this paper focuses on the imaging spectroscopy of the spectral bump. Thus, we focus on the spectral and spatial characteristics of the type II burst in the time range 11:25:00\,UT to 11:30:30\,UT.
We categorize this 5.5-minute time range into three phases: P1, P2, and P3, representing before, during, and after the bump, respectively. As shown in Figure \ref{fig:main-ds}.

The type II shows a band-splitting into two bands with a $\sim$8~MHz frequency separation. The band split lanes are determined by finding the local maxima on the upper and lower split bands respectively. Fig \ref{fig:main-ds} upper panels present the FWHM and peak location of the brightness temperature distribution of the higher-frequency band (HB; blue) and the lower-frequency band (LB; green), for three time slots P1, P2, and P3 labeled in the dynamic spectrum.

\begin{figure}[h!]
    \centering
    \includegraphics[width=0.44\textwidth]{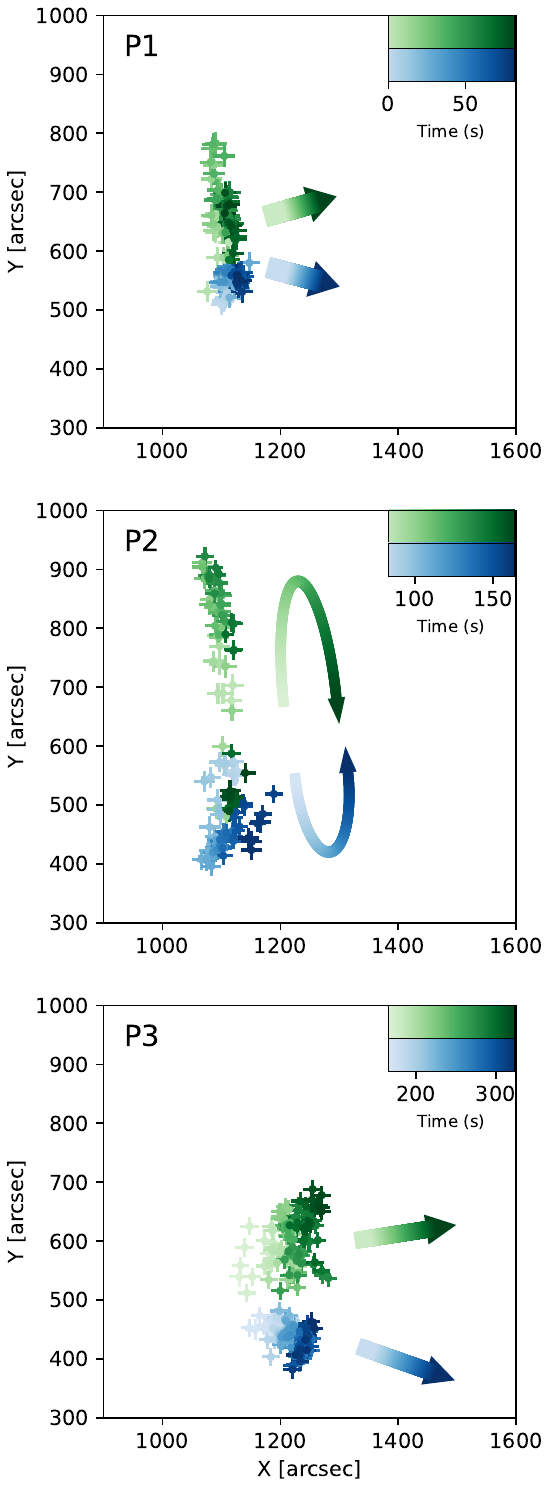}
    \caption{The source location of the type II burst during the three phases, the points are colored according to time. The arrows indicate the source location variation trend.}
    \label{fig:interf3phase}
\end{figure}

\section{Results}

From the imaging along the type II bands, we can see that, the source movement of type II shows different behavior during the P1, P2, and P3 phases, as shown in Figure \ref{fig:interf3phase}.
The HB (blue) and LB (green) radio sources are separated in space vertically. During P1 and P3, the peak locations of the HB and LB sources both move from east to west, which indicates that the shock is moving from east to west as well, as shown by the blue and the green arrow in panel P1 and P3 of Fig \ref{fig:interf3phase}. However, during P2 (the time of spectral bump), \peijin{the upper and lower sources sharply separated in the north-south direction and stayed separated for $\sim$1 minute and then converged back to a closer distance toward the end of the spectrum bump.}

\begin{figure}
    \centering
    \includegraphics[width=0.46\textwidth]{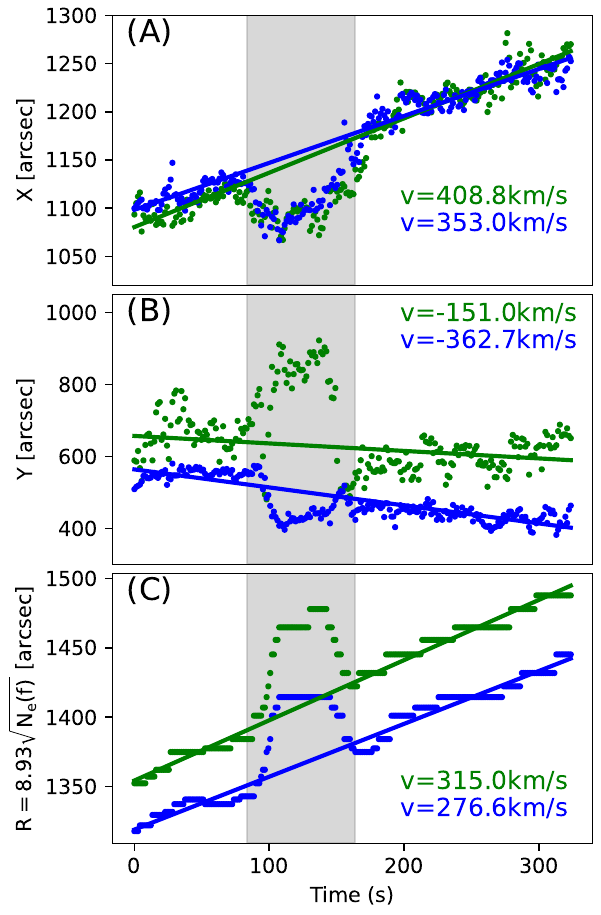}
    \caption{The source location variation in X and Y directions are shown in panel A and panel B. The heliocentric distance of the source derived from the frequency of the imaging is shown in panel C. The solid lines are the fitted result of the points in Phase-1 and Phase-3 (i.e. the points out of the gray area).}
    \label{fig:speed}
\end{figure}

By fitting the source location variation with time, we measured its speed projected in the sky plane.
As shown in Figure \ref{fig:speed}
During P1 and P3, (According to Panel A and B in Figure \ref{fig:speed}) the HB and LB sources have speeds in the x-direction of 353 and 409\,km/s, respectively, and y-direction speeds of -363 and -151\,km/s, respectively. This indicates that the radio source motion is mostly from east to west, and partially to the south. With an average speed to the west direction of 381 km/s, and 76 seconds duration of P2, the estimated width of the density depletion is 29Mm.  
During the spectral bump (phase P2), the source of both HB and LB has a negative deviation from the movement trend of P1 and P3. In the x-direction, this deviation is to the west. In the y-direction, the source deviates towards the north for LB and towards the south for the HB, \peijin{and the source of LB and HB converge back to the middle toward the end of the spectrum bump }.

\zpeijin{We estimated the heliocentric distance of the radio source from the frequency of the emission. To compare with the source location from the imaging of the harmonic component in this type II, we assume harmonic emission in the estimation ($f=2\omega _{\mathrm {pe} }$ ), where  $\omega _{\mathrm {pe} }$ is the plasma frequency expressed as:}
\zpeijin{
$${\displaystyle \omega _{\mathrm {pe} }={\sqrt {\frac {N_{\mathrm {e}}(R) \, e^{2}}{m_{e}\varepsilon _{0}}}}}. $$ 
 $N_e(R)$ is background electron density from the empirical model  \citet{saito77}. }  
The result is shown in Figure \ref{fig:speed} Panel (C), where the higher and lower lanes have an estimated radial speed of 276 and 315 km/s from P1 and P3. The speed estimated from the empirical density model is slower by $\sim$100~km/s than the source motion speed obtained from imaging.


\begin{figure}
    \centering
    \includegraphics[width=0.49\textwidth]{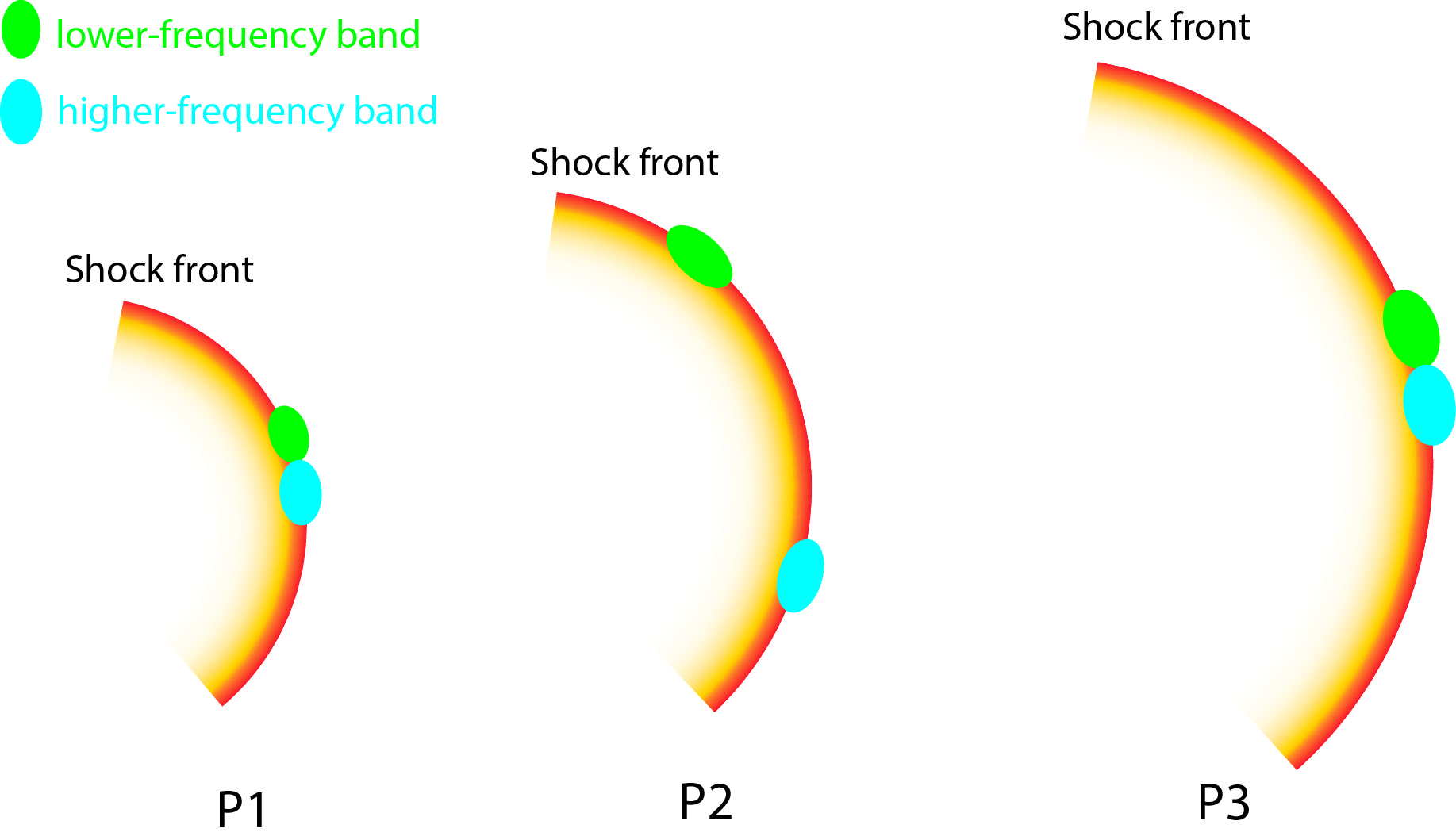}
    \caption{Shock front evolution and relative source location, red arc represents the shock front, the blue and green region represents the location of radio source}
    \label{fig:catoon}
\end{figure}

\section{Discussion and conclusion}

The observations and results of this study can be summarised into the following points:
{\begin{enumerate}
    \item The `spectral bump' in the type II radio burst is 8~MHz at 60~MHz (for harmonic wave) corresponding to a density drop of $\sim$25\%.
    \item Overall source moving trend is to the east direction (corresponds to the shock preceding direction), the speed in the east-west direction is 408~km/s and 353~km/s measured from the upper and lower lanes.
    \item The distance the shock crosses through the time period of the spectral bump is 29Mm, which is too small to identify at other wavelengths. There are coronal holes near the west limb, and also there are density fluctuations in the density model but no direct counterpart can be found.
    \item At the peak of the bump, the source separation is $\sim$500 arcseconds into the northwest direction.
\end{enumerate}}


Structures in Type II radio bursts can reflect structures in the solar corona. 
The spectral bump analyzed in this work indicates that the shock encounters a low density compact region. We investigated extreme-ultraviolet (EUV) images to identify if there is a possible coronal hole or lower density region in the direction of the type II propagation. There is indeed a small coronal hole with sharp boundaries rotating onto the west limb in images from the Atmospheric Imaging Assembly onboard the Solar Dynamic Observatory \citep{lemen2012}. This coronal hole is clearly visible two to three days before the type II burst. However, it is hard to identify these coronal holes during the present observations as they are located too close to the solar limb. It is possible that the low density and open magnetic field of a coronal hole can cause a spectral bump accompanied by a source separation when encountered by the shock. A cartoon of the source deviation is shown in Figure \ref{fig:catoon}. In this cartoon, the movement in the x-direction can be explained by the source being emitted at further locations along a curved shock front. The separation in y-direction is likely due to the shock encountering a density drop (such as a coronal hole boundary) causing either the emission to shift or to fan outwards with the open coronal magnetic field lines originating from a coronal hole.


{The observed spatial separation between the two bands also confirms the theory where band-splitting is caused by emission at different locations upstream of the shock as initially suggested by \citet{holman1983}. This is in particular very clear during the spectral bump, where such a high separation ($\sim$500~arcsec which corresponds to 0.5~R$_\odot$) cannot be attributed to radio propagation effects. The radio source location can be affected by propagation effects such as scattering and refraction. These can cause a 0.2 $R_{\rm sun}$ radial offset for fundamental wave and 0.05 $R_{\rm sun}$ for harmonic wave for a 35MHz radio source near limb \citep{propagationZhang2021ApJ}. In the present study we only consider the harmonic emission of a type II at even higher frequencies and find a large separation of 0.5~R$_\odot$. Thus, propagation effects are most likely negligible in our study. More recent studies have also been in agreement with the \citet{holman1983} theory of band-split emission, finding consistent separations of the split-bands throughout the evolution of the type II similar to the present study \citep[e.g.,][]{bhunia2022imagingspectroscopy, morosan2023type}.}



%


\begin{acknowledgements}

P.Z. and D.E.M. acknowledge the University of Helsinki Three-Year Grant. 
P.Z. is supported by the NASA Living With a Star Jack Eddy Postdoctoral Fellowship Program, administered by UCAR's Cooperative 
Programs for the Advancement of Earth System Science (CPAESS) under award NNX16AK22G. 
D.E.M. and S.N. acknowledge the Research Council of Finland project `SolShocks' (grant number: 354409). D.E.M also acknowledges the Academy of Finland project `RadioCME' (grant number 333859).
C. V. has been supported by the Deutsche Forschungsgemeinschaft (DFG,
German Research Foundation) under project number VO 2123/1-1
The authors wish to acknowledge CSC – IT Center for Science, Finland, for computational resources.
LOFAR \citep{van2013lofar} is the Low Frequency Array designed and constructed by ASTRON. Data available at \href{https://lta.lofar.eu/}{https://lta.lofar.eu/}. The LOFAR ILT resources have benefited from the following recent major funding sources: CNRS-INSU, Observatoire de Paris and Universite Orleans, France; BMBF, MIWF-NRW, MPG, Germany; Science Foundation Ireland (SFI), Department of Business, Enterprise and Innovation (DBEI), Ireland; NWO, The Netherlands; The Science and Technology Facilities Council, UK; The Ministry of Science and Higher Education (MEiN), Poland. 
UWM thanks the National Science Centre, Poland for granting “LOFAR
observations of the solar corona during Parker Solar Probe perihelion
passages” in Beethoven Classic 3 funding initiative under project number 2018/31/G/ST9/01341. 
UWM would also like to thank (MEiN) for its contribution to the ILT, LOFAR2.0 upgrade (decision numbers: 2021/WK/2, 30/530252/SPUB/SP/2022, 29/530358/SPUB/SP/2022, 28/530020/SPUB/SP/2022, respectively).

\end{acknowledgements}

%
\bibliographystyle{aa} 
\bibliography{cite} 
%

\end{document}